\documentclass{phbauth}        % Don't modify
\usepackage{graphicx,epsf}

\begin{document}

\begin{frontmatter}

% Use lower case letters in the title.
\title{Finite H$_2$ concentrations in superfluid $^4$He}

\author[]{J. M. Mar\'\i n},
\author[]{J. Boronat\thanksref{thank1}},
\author[]{J. Casulleras}

\address{Departament de F\'\i sica i Enginyeria Nuclear, Campus
Nord B4-B5, Universitat Polit\`ecnica de Catalunya, \protect \\ E-08034 
Barcelona, Spain}

% The corresponding author should be distinguished and his email
% address and/or fax number must be given. His mailing address has to
% be complete: the proofs are send to this address around
% January 1, 2000. The address for sending proofs has to be indicated
% as "present address", if it is different from the address above.
\thanks[thank1]{Corresponding author. E-mail: boronat @ fen.upc.es} 

\begin{abstract}
We have studied the solubility of molecular hydrogen in bulk
liquid $^4$He at zero temperature using the diffusion Monte Carlo method
and realistic interatomic potentials between the different species of the
mixture.
Around the $^4$He equilibrium density, the H$_2$ molecules clusterize in
liquid-like drops blocking the existence of a uniform dilution. On the
contrary,
at higher densities the cluster formation is less feasible and
metastable dilute solutions may exist. 
\end{abstract}


\end{frontmatter}

% The main text begins here. The \section commands are optional.
%\section{Introduction}

The achievement of a stable liquid phase of molecular hydrogen H$_2$ that
might exhibit superfluid behaviour has been object of a continued
experimental and theoretical interest. Bulk H$_2$ is a hcp solid that melts
at $13.8$ K whereas the critical temperature of its superfluid transition
in bulk is expected to be much smaller (3-4 K) \cite{silvera}. This difference
 has proved
too large to be experimentally covered by supercooling, even in confined
geometries.

We study the liquid $^4$He-H$_2$ mixture at zero temperature by means of 
diffusion Monte Carlo (DMC). This method solves stochastically the $N$-body
Schr\"odinger equation providing results which are exact for bosonic
systems like the present one. The hydrogen molecules are considered in
their ground state (para-H$_2$) and interact by means of the isotropic
potential proposed by Silvera and Goldman \cite{goldman}. The 
potential of Meyer {\it et al} \cite{meyer} and the 
HFD-B(HE) Aziz potential \cite{aziz} have been used for the
$^4$He-H$_2$ and  $^4$He-$^4$He pair
interactions, respectively. In all the calculations we have used a simulation box
containing 108 particles, 5 of them being H$_2$ molecules, what suppose a
macroscopic H$_2$ concentration of $\sim 5$\%.

The dilution of H$_2$ in liquid $^4$He could be favoured by its lower mass,
roughly a factor two, that increases its kinetic energy with respect
$^4$He. However, the well of the H$_2$-H$_2$ interaction is three times
deeper than the one of the He-He potential, a fact that points to a
clusterization of the H$_2$ molecules in order to minimize the energy. Our
results show that the latter effect dominates in a wide range of densities.
The $^4$He-H$_2$ mixture has been studied at total densities $0.328$,
$0.365$, $0.401$, and $0.424\ \sigma^{-3}$ ($\sigma=2.556$ \AA), and in
each case, two different types of initial configurations have been chosen.
In the first type, the H$_2$ molecules are uniformly distributed in the simulation
box, while in the second one, they are all close one another in a
cluster-like structure. 

A common trend to all the densities is that the
cluster structure is energetically preferred to the uniform mixture but
this difference reduces when the density increases. The
lifetime of the metastable homogeneous solution increases when that energy
difference decreases, and therefore, a uniform mixture is more feasible at
higher densities.
The DMC results confirm this behaviour: 
the simulations  starting on
a uniform dilution evolve towards a molecular aggregate in 
the scale of a typical run. Once the cluster is formed, its structural evolution
and stability depends on the density. 
\begin{figure}[t]
\begin{center}
\leavevmode
\includegraphics[width=0.9\linewidth]{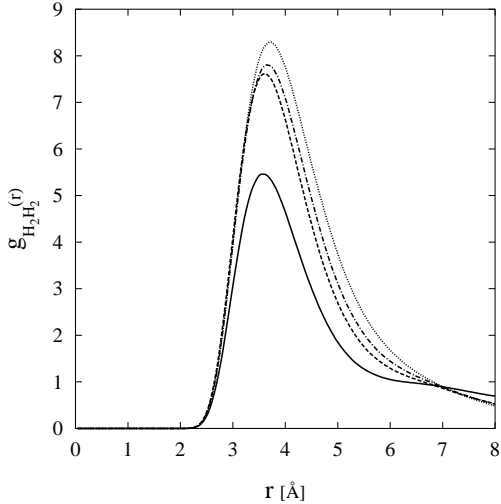}
\caption{
H$_2$-H$_2$ radial distribution function for densities $0.328$,
$0.365$, $0.401$, and $0.424\ \sigma^{-3}$, from top to bottom.}
\end{center}
\label{figur1}
\end{figure}                                                                         
\begin{figure}[t]
\begin{center}
\leavevmode
\includegraphics[width=0.9\linewidth]{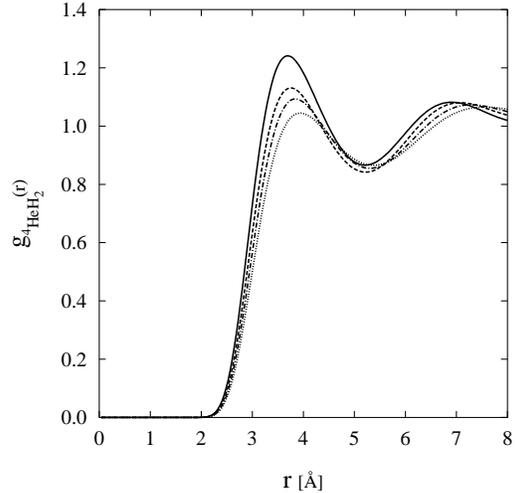}
\caption{
$^4$He-H$_2$ radial distribution function for densities $0.328$,
$0.365$, $0.401$, and $0.424\ \sigma^{-3}$, from bottom  to top.}
\end{center}
\label{figur2}
\end{figure}                                                                         

The two-body radial distribution function $g_{{\rm H}_2-{\rm H}_2}(r)$ is
shown in Fig. 1 for the four densities analyzed. In all cases, the
functions plotted correspond to the stable regime once the initial dilute
regime has disappeared. The different behaviour between the first three 
densities and the highest one is reflected in the figure. For densities
below $0.424\ \sigma^{-3}$ the three distribution functions are very
similar, with a high central peak and a tail that approaches zero at medium
distance. This structure points unambiguously to the existence of a liquid
H$_2$ cluster moving in the $^4$He medium. The curve at $0.424\
\sigma^{-3}$ presents a lower central peak and values closer to one at
large distances suggesting a coexistence of dilute and
cluster structures.

The crossed distribution function $g_{^4{\rm He}-{\rm H}_2}(r)$ provides
additional information on the local structure around the H$_2$ molecules. In
Fig. 2, results for $g_{^4{\rm He}-{\rm H}_2}(r)$ at different densities
are reported. In this case, the stability of the cluster structure produces
a depression of the height of the first peak and a subsequent increase of
the second peak that, at low densities, is even higher than the first one.
On the contrary, at $0.424\ \sigma^{-3}$ one observes a behaviour much 
closer to the characteristic one in a homogeneous mixture.

In conclusion, only at high densities and low H$_2$ concentrations a
metastable uniform $^4$He-H$_2$ mixture may exist. Below a characteristic
density ($\sim 0.424\ \sigma^{-3}$) a liquid cluster of H$_2$ is
predominantly  observed.

%\section{xxx}

% Acknowledgements are optional.
%\begin{ack}
%xxx
%\end{ack}

% References

\end{document}